\documentclass[twocolumn]{aastex631}
\usepackage{enumitem}
\usepackage{soul}
\usepackage{amsmath}
\usepackage{bm}

\accepted{for Publication in ApJ}

\shorttitle{HETDEX Survey for the $\mathcal{M}_*-\mathcal{M}_\mathrm{BH}$ Relation Evolution}
\shortauthors{Zhang et al.}

\begin{document}

\title{The Stellar Mass - Black Hole Mass Relation at $z\sim2$ Down to $\mathcal{M}_\mathrm{BH}\sim10^7 M_\odot$ Determined by HETDEX}

\correspondingauthor{Yechi Zhang}
\email{yczhang@icrr.u-tokyo.ac.jp}

\author[0000-0003-3817-8739]{Yechi Zhang}
\affiliation{Institute for Cosmic Ray Research, The University of Tokyo, 5-1-5 Kashiwanoha, Kashiwa, Chiba 277-8582, Japan}
\affiliation{Department of Astronomy, Graduate School of Science, the University of Tokyo, 7-3-1 Hongo, Bunkyo, Tokyo 113-0033, Japan}
\affiliation{Kavli Institute for the Physics and Mathematics of the Universe (Kavli IPMU, WPI), The University of Tokyo, 5-1-5 Kashiwanoha, Kashiwa, Chiba, 277-8583, Japan}

\author[0000-0002-1049-6658]{Masami Ouchi}
\affiliation{National Astronomical Observatory of Japan, 2-21-1 Osawa, Mitaka, Tokyo 181-8588, Japan}
\affiliation{Institute for Cosmic Ray Research, The University of Tokyo, 5-1-5 Kashiwanoha, Kashiwa, Chiba 277-8582, Japan}
\affiliation{Kavli Institute for the Physics and Mathematics of the Universe (Kavli IPMU, WPI), The University of Tokyo, 5-1-5 Kashiwanoha, Kashiwa, Chiba, 277-8583, Japan}

\author[0000-0002-8433-8185]{Karl Gebhardt}
\affiliation{Department of Astronomy, The University of Texas at Austin, 2515 Speedway Boulevard, Austin, TX 78712, USA}

\author[0000-0001-5561-2010]{Chenxu Liu}
\affiliation{South-Western Institute for Astronomy Research, Yunnan University, Kunming, Yunnan, 650500, People’s Republic of China}
\affiliation{Department of Astronomy, The University of Texas at Austin, 2515 Speedway Boulevard, Austin, TX 78712, USA}

\author[0000-0002-6047-430X]{Yuichi Harikane}
\affiliation{Institute for Cosmic Ray Research, The University of Tokyo, 5-1-5 Kashiwanoha, Kashiwa, Chiba 277-8582, Japan}

\author[0000-0002-2307-0146]{Erin Mentuch Cooper}
\affiliation{Department of Astronomy, The University of Texas at Austin, 2515 Speedway Boulevard, Austin, TX 78712, USA}
\affiliation{McDonald Observatory, The University of Texas at Austin, 2515 Speedway Boulevard, Austin, TX 78712, USA}

\author[0000-0002-8925-9769]{Dustin Davis}
\affiliation{Department of Astronomy, The University of Texas at Austin, 2515 Speedway Boulevard, Austin, TX 78712, USA}

\author[0000-0003-2575-0652]{Daniel J. Farrow}
\affiliation{University Observatory, Fakult\"at f\"ur Physik, Ludwig-Maximilians University Munich, Scheinerstrasse 1, 81679 Munich, Germany}
\affiliation{Max-Planck Institut f\"ur extraterrestrische Physik, Giessenbachstrasse 1, 85748 Garching, Germany}

\author[0000-0003-1530-8713]{Eric Gawiser}
\affiliation{Physics and Astronomy Department, Rutgers, The State University of New Jersey, Piscataway, NJ 08854, USA}

\author[0000-0001-6717-7685]{Gary J. Hill}
\affiliation{McDonald Observatory, The University of Texas at Austin, 2515 Speedway Boulevard, Austin, TX 78712, USA}
\affiliation{Department of Astronomy, The University of Texas at Austin, 2515 Speedway Boulevard, Austin, TX 78712, USA}

\author[0000-0002-0417-1494]{Wolfram Kollatschny}
\affiliation{Institut f\"ur Astrophysik und Geophysik, Universit\"at G\"ottingen, Friedrich-Hund Platz 1, 37077 G\"ottingen, Germany}

\author[0000-0001-9011-7605]{Yoshiaki Ono}
\affiliation{Institute for Cosmic Ray Research, The University of Tokyo, 5-1-5 Kashiwanoha, Kashiwa, Chiba 277-8582, Japan}

\author[0000-0001-7240-7449]{Donald P. Schneider}
\affiliation{Department of Astronomy \& Astrophysics, The Pennsylvania State University, University Park, PA 16802, USA}
\affiliation{Institute for Gravitation and the Cosmos, The Pennsylvania State University, University Park, PA 16802, USA}


\author[0000-0001-8519-1130]{Steven L. Finkelstein}
\affiliation{Department of Astronomy, The University of Texas at Austin, 2515 Speedway Boulevard, Austin, TX 78712, USA}

\author[0000-0001-6842-2371]{Caryl Gronwall}
\affiliation{Department of Astronomy \& Astrophysics, The Pennsylvania
State University, University Park, PA 16802, USA}
\affiliation{Institute for Gravitation and the Cosmos, The Pennsylvania State University, University Park, PA 16802, USA}

\author[0000-0002-1590-0568]{Shardha Jogee}
\affiliation{Department of Astronomy, The University of Texas at Austin, 2515 Speedway Boulevard, Austin, TX 78712, USA}

\author{Mirko Krumpe}
\affiliation{Leiniz-Institut f\"ur Astrophysik Potsdam, An der Sternwarte 16, 14882 Potsdam, Germany}

\begin{abstract}
We investigate the stellar mass - black hole mass ($\mathcal{M}_*-\mathcal{M}_\mathrm{BH}$) relation with type 1 AGN down to $\mathcal{M}_\mathrm{BH}=10^7 M_\odot$, corresponding to a $\simeq -21$ absolute magnitude in rest-frame ultraviolet (UV), at $z = 2-2.5$. Exploiting the deep and large-area spectroscopic survey of the Hobby-Eberly Telescope Dark Energy Experiment (HETDEX), we identify 66 type 1 AGN with $\mathcal{M}_\mathrm{BH}$ ranging from $10^7$ to $10^{10} M_\odot$ that are measured with single-epoch virial method using C{\sc iv} emission lines detected in the HETDEX spectra. $\mathcal{M}_*$ of the host galaxies are estimated from optical to near-infrared photometric data taken with Spitzer, WISE, and ground-based 4-8m class telescopes by \texttt{CIGALE} SED fitting. We further assess the validity of SED fitting in two cases by host-nuclear decomposition performed through surface brightness profile fitting on spatially-resolved host galaxies with JWST/NIRCam CEERS data. We obtain the $\mathcal{M}_*-\mathcal{M}_\mathrm{BH}$ relation covering the unexplored low-mass ranges of $\mathcal{M}_\mathrm{BH}~\sim~10^7-10^8~M_\odot$, and conduct forward modelling to fully account for the selection biases and observational uncertainties. The intrinsic $\mathcal{M}_*-\mathcal{M}_\mathrm{BH}$ relation at $z\sim 2$ has a moderate positive offset of $0.52\pm0.14$~dex from the local relation, suggestive of more efficient black hole growth at higher redshift even in the low-mass regime of $\mathcal{M}_\mathrm{BH}~\sim~10^7-10^8~M_\odot$. Our $\mathcal{M}_*-\mathcal{M}_\mathrm{BH}$ relation is inconsistent with the $\mathcal{M}_\mathrm{BH}$ suppression at the low-$\mathcal{M}_*$ regime predicted by recent hydrodynamic simulations at a $98\%$ confidence level, suggesting that feedback in the low-mass systems may be weaker than those produced in hydrodynamic simulations.
\end{abstract}


\section{Introduction} \label{sec:intro}

It has been known for over two decades that in the local universe, the mass of supermassive black holes (SMBHs) are tightly correlated with the bulge properties (e.g., velocity dispersion, bulge mass) of their host galaxies  \citep[e.g.,][]{magorrian98,gebhardt00,kh13}, indicating the strong connection between the growth of central SMBHs and their host galaxies. Although hydrodynamic simulations have successfully reproduced the strong correlation between the growth of SMBH and their host galaxy, the underlying physical mechanisms are still under debate. One of the scenarios is that central SMBHs regulate the host galaxies, or vise versa, by various kinds of feedback effects \citep[e.g.,][]{springel05a,hopkins08a} or sharing the same gas reservoir \citep[e.g.,][]{menci16,ni21}. Conversely, it has been proposed that galaxy assembly processes, such as mergers, can produce a strong observed correlation without the necessity of the physical coupling between SMBHs and their host galaxies \citep[e.g.,][]{hirschmann10,jahnke11}.

Previous galactic evolution studies have focused on active galactic nuclei (AGN), or quasars, at high redshift and measured their black hole (BH) mass ($\mathcal{M}_\mathrm{BH}$) and total stellar mass ($\mathcal{M}_*$). For example, \citet{schramm13} and \citet{mechtley16} decomposed the host and AGN emission and derived the $\mathcal{M}_*-\mathcal{M}_\mathrm{BH}$ relation out to $z=1-2$, reporting that there is no evolution compared with the local relation.  
\citet{merloni10} collected quasars at $z\sim 2$ and derived their $\mathcal{M}_*$ by decomposing the spectral energy distributions (SEDs) into a stellar component and an AGN component, finding that the majority of their objects are located above the local $\mathcal{M}_*-\mathcal{M}_\mathrm{BH}$ relation, suggesting that AGN at earlier epochs tend to host overmassive BHs. Such a trend was also discussed by \citet{ding20}, who reported a mildly positive evolution in $\mathcal{M}_\mathrm{BH}/\mathcal{M}_*$ ratio with increasing  redshift out to $z\sim 1.7$, although the non-evolution scenario is also plausible at the $2\sigma-3\sigma$ confidence level.

Another open question on the $\mathcal{M}_*-\mathcal{M}_\mathrm{BH}$ relation is whether this linear relation holds at the low-mass end. Some hydrodynamic simulations have predicted that below a critical $\mathcal{M}_*$, the $\mathcal{M}_*-\mathcal{M}_\mathrm{BH}$ relation would deviate from a linear shape, characterized by a population with undermassive BHs due to the smaller $\mathcal{M}_\mathrm{BH}$ of seed BHs \citep[e.g.,][]{yajima22} or from strong supernova feedback \citep[e.g.,][]{sijacki15}. However, observations of high-z AGN have not yet reached the low-mass regime required to distinguish whether such a break in $\mathcal{M}_*-\mathcal{M}_\mathrm{BH}$ relation exists or not.

To target these open questions, it is important to push the observations of high-redshift AGN towards the low-mass regime. In this study, we extend observations of $\mathcal{M}_*-\mathcal{M}_\mathrm{BH}$ to low masses using the untargeted integral field spectroscopic survey of the Hobby-Eberly Telescope Dark Energy Experiment \citep[HETDEX,][]{gebhardt21,hill21}. HETDEX utilizes the VIRUS wide-field spectrograph \citep{hill21} on the upgraded 10 m Hobby-Eberly Telescope \citep[HET,][]{ramsey98, hill21}. AGN detected in the HETDEX survey include faint type 1 AGN that potentially host SMBHs with low $\mathcal{M}_\mathrm{BH}$ \citep{liu22a,liu22b,zhang21}. Utilizing multiband archival photometry in the Spitzer-HETDEX Exploratory Large-area (SHELA) Survey field \citep{papovich16}, we perform host-nuclear decomposition with SED fitting and derive $\mathcal{M}_*$. Measurement uncertainties are estimated by conducting Monte Carlo simulations and comparing with the latest NIRCam \citep{rieke05} imaging data of the early release science (ERS) of the James Webb Space Telescope (JWST). With the forward modelling that incorporates the selection functions and measurement uncertainties in $\mathcal{M}_*$ and $\mathcal{M}_\mathrm{BH}$, we investigate the intrinsic $\mathcal{M}_*-\mathcal{M}_\mathrm{BH}$ relation at $z=2.0-2.5$ down to $\log(\mathcal{M}_\mathrm{BH}/\mathcal{M}_\odot)~\sim~7$, a mass regime that has not yet been explored before at such a redshift range.

This paper is organized as follows. Section \ref{sec:data} introduces our HETDEX type 1 AGN sample and the ancillary photometric data. Section \ref{sec:methods} describes the $\mathcal{M}_*$ and $\mathcal{M}_\mathrm{BH}$ measurements and the potential systematics. The observed $\mathcal{M}_*-\mathcal{M}_\mathrm{BH}$ relation is presented in Section \ref{sec:obs_results}. The forward modelling that accounts for selection biases and measurement uncertainties is presented in Section \ref{sec:model}, the intrinsic $\mathcal{M}_*-\mathcal{M}_\mathrm{BH}$ relation is derived in Section \ref{sec:int_results}; discussion of the results follows in Section \ref{sec:discussion}. Throughout this paper, we use AB magnitudes \citep{oke74} and the cosmological parameters of ($\Omega_\mathrm{m}$, $\Omega_k$, $H_0$) = (0.315, 0.001, 67.4~km~s$^{-1}$~Mpc$^{-1}$) according to \citet{planck18}.

\section{Data and Sample Selection} \label{sec:data}
\subsection{HETDEX type~1 AGN}
The AGN sample used in this study is selected from the HETDEX AGN catalog \citep{liu22a}. Here we briefly describe the HETDEX AGN catalog, and refer the readers to \citet{liu22a} for full details. 
With the fiber spectral data of the HETDEX survey that cover the wavelength range of $3500-5500$~\AA \citep{gebhardt21}, they performed the emission line detection with a grid search and identified 2346 AGN. The emission line properties of each AGN are obtained by fitting the AGN spectra with a power-law continuum at the selected continuum windows and multiple Gaussian profiles at the expected wavelengths of each detected emission lines. These 2346 type~1 AGN spans a redshift range of 1.88-3.53. The UV luminosity function of the HETDEX type~1 AGN is comparable with previous results at similar redshifts while extending to a rest-frame UV continuum magnitude of $-21$, three magnitudes fainter than previous results \citep{zhang21,liu22b}. 

From the HETDEX type 1 AGN catalog presented in \citet{liu22a}, we select objects that: a) are located in the SHELA field, and b) have redshifts of $z=2.0-2.5$ and broad (FWHM$> 1000$~km~s$^{-1}$) C{\sc iv} emission lines with emission line fluxes greater than $2\times10^{-16}$~erg~s$^{-1}$~cm$^2$ that correspond to the 50\% detection limit of typical broad emission lines in the HETDEX spectra \citep{liu22b}. Because C{\sc iv} emission line profiles of type~1 AGN may show absorption features and affect the line profile fitting, we further require the fitting results of C{\sc iv} emission lines to have chi2 values smaller than four and conduct visual inspection to exclude the objects with bad fitting results. The total number of type 1 AGN selected is 77. 


\subsection{Multi-band Photometry}
\label{subsec:phot}
We collect ancillary archival multi-band photometric data, including $grizy$-band imaging from the third data release (DR3) of the Subaru/HyperSupreme Cam Strategic Survey Program \citep[HSC-SSP;][]{aihara18,aihara22}, $J$-band imaging from the VISTA/VICS82 survey \citep{geach17}, $K_s$-band imaging from the KPNO/NEWFIRM HETDEX Survey \citep[NHS;][]{stevans21}, 3.6 and 4.5~$\mu$m Spitzer-IRAC imaging from the SHELA survey \citep{papovich16}, as well as 12 and 22~$\mu$m imaging from the AllWISE survey \citep{wright10,cutri13}. These photometric data cover an observed wavelength of 0.4-20~$\mu$m. The list of photometric data used in this study is summarized in Table \ref{tab:multiband}.
\begin{deluxetable}{cccc}[htbp!]
\tablecaption{Summary of the multi-band data used for the SED fitting analysis.
\label{tab:multiband}}
\tablecolumns{4}
\tablewidth{0pt}
\tablehead{
\colhead{Telescope/Instrument} & \colhead{Filter} & \colhead{$\lambda_\mathrm{eff}$ \tablenotemark{a}} & \colhead{5$\sigma$ depth} \\
& & \colhead{$\mu$m} & \colhead{AB mag}
}
\startdata
Subaru/HSC & $g$ & 0.479 & 26.5 \\
 & $r$ & 0.619 & 26.5 \\
 & $i$ & 0.767 & 26.2 \\
 & $z$ & 0.89 & 25.2 \\
 & $y$ & 0.978 & 24.4 \\ \hline
VISTA/VIRCAM & $J$ & 1.252 & 20.9 \\ \hline
 KPNO/NEWFIRM & $K_s$ & 2.152 & 22.4 \\ \hline
\textit{Spitzer}/IRAC & $ch1$ & 3.556 & 22.0 \\
 & $ch2$ & 4.501 & 22.0 \\ \hline
\textit{WISE} & 12~$\mu m$ & 11.561 & 14.1 \\
 & 22~$\mu m$ & 22.088 & 14.5 \\
\enddata
\tablenotetext{a}{Effective wavelength}
\end{deluxetable}

\subsubsection{HSC-SSP DR3}
The HSC-SSP DR3 \citep{aihara22} includes deep multi-band imaging data covering a sky area of 670~deg$^2$. The typical seeing sizes for the five broadband filters ($g$, $r$, $i$, $z$, $y$) are $0.\!\!\arcsec6 - 0.\!\!\arcsec8$. The data reduction and source detection are performed with \texttt{hscPipe v6.7} \citep{bosch18}. From the HSC database\footnote{\url{https://hsc-release.mtk.nao.ac.jp/doc/}} we cross-match our HETDEX type 1 AGN sample to the HSC-SSP DR3 multi-band catalog within $2''$ radii, and adopt the following criteria in all of the five broadband filters to remove spurious sources:
\begin{enumerate}[noitemsep]
  \item isPrimary = True
  \item nchild = 0 
  \item pixelflags\_edge = False
  \item pixelflags\_interpolatedcenter = False
  \item pixelflags\_saturatedcenter = False
  \item pixelflags\_crcenter = False
  \item pixelflags\_bad = False
  \item pixelflags\_bright\_objectcenter = False
  \item pixelflags\_bright\_object = False
  \item pixelflags = False
\end{enumerate}
All of our HETDEX type 1 AGN are detected in at least one band in HSC-SSP images. We use the Kron magnitudes reported in the HSC-SSP DR3 catalog as the total continuum flux densities.

\subsubsection{VICS82/NHS/SHELA}
We collect the near-IR photometry from the multi-band catalog of \citet{stevans21}. Based on their NHS $K_s$-band imaging taken with the KPNO Mayall 4~m Telescope, \citet{stevans21} constructed the $K_s$-band source catalog and merged with the VISTA $JK_s$-band catalog from the VICS82 survey \citep{geach17} and the Spitzer/IRAC 3.6, 4.5~$\mu$m catalog from the SHELA survey \citep{papovich16}. Since the limiting magnitude of NHS $K_s$-band photometry (5$\sigma$ 22.4 mag in a $2''$-diameter aperture) is fainter than that of VICS82 survey (5$\sigma$ 20.9 mag in a $2''$-diameter aperture), we use the NHS imaging for the $Ks$-band photometry. We cross-match our HETDEX type 1 AGN sample to the \citet{stevans21} multi-band catalog within $2''$ radii. All of type 1 AGN are matched with NHS $K_\mathrm{s}$-band detected objects. We use the "AUTO" fluxes in the NHS catalog, which is equivalent to the Kron fluxes, as the total continuum flux densities.

\subsubsection{AllWISE}
The AllWISE survey, based on the Wide-field Infrared Survey Explorer mission \citep[WISE;][]{wright10}, provides all-sky mid-infrared imaging with four bands at 3.4, 4.6, 12, and 22~$\mu$m (hereafter $W1$, $W2$, $W3$, and $W4$-bands). We cross-match the HETDEX type 1 AGN sample to the AllWISE catalog \footnote{The ALLWISE source catalog is available at IPAC: \dataset[10.26131/irsa1]{\doi{10.26131/irsa1}}
}with a matching radius of $2''$. We convert the Vega magnitudes ($m_\mathrm{Vega}$) reported in the AllWISE catalog to the AB magnitudes ($m_\mathrm{AB}$) by $m_\mathrm{AB} = m_\mathrm{Vega} + \Delta m$, where $\Delta m = (2.699, 3.339, 5.174, 6.620)$ for the $W1$, $W2$, $W3$, and $W4$-bands, respectively \citep{cutri13}. Because the $W1$ and $W2$-bands have the similar wavelength coverage as the $Spitzer$/IRAC $ch1$ and $ch2$ while having shallower limiting magnitudes, we only take the $W3$ and $W4$-bands photometry for the SED fitting. Out of the 77 HETDEX AGN, nine objects have detections in AllWISE $W3$ and/or $W4$ photometry while the other 68 objects have no detections in either $W3$ or $W4$ bands. For non-detections, we apply the 1$\sigma$ flux limit of ALLWISE \citep[Table \ref{tab:multiband}, also see][]{wright10}.

\section{Methods} \label{sec:methods}

\subsection{Host-nuclear decomposition and $M_*$ measurements}
\label{subsec:mstar}

We conduct SED fitting to decompose the nuclear and stellar components of our type 1 AGN with the code \texttt{CIGALE} \citep{boq19},
which models the SEDs of stellar, nebular, dust, and AGN components in a self-consistent way by considering the energy balance between the UV/optical and IR. For the stellar continuum component, we adopt the stellar population synthesis models of \citet{bc03} with star formation histories of the single exponential decreasing star formation assuming the initial mass function of \citet{chabrier03}. The dust component is modelled with the dust-emission template of \citet{dale14} and the dust attenuation law of \citet{calzetti00}. The nebular emission is calculated with the template of \citet{inoue11}. For the AGN component, we apply the \citet{fritz06} model that describes the structure and geometry of the dusty torus and calculates the radiation transfer. The modeled SEDs are then redshifted, and attenuation by the intergalactic medium (IGM) is considered with the prescription of \citet{meiksin06}. The redshift of each object is fixed to the spectroscopic redshift measured with the HETDEX spectra. We restrict the ranges of free parameters in the models as summarized in Table \ref{tab:CIGALE}, referring to previous studies that use the {\tt CIGALE} code \citep{yang18, boq19, li21b}. Specifically, the input parameters of AGN emission are taken from the typical type 1 AGN template from \citet{ciesla15}.

\begin{figure}[ht!]
\begin{center}
\includegraphics[scale=0.41]{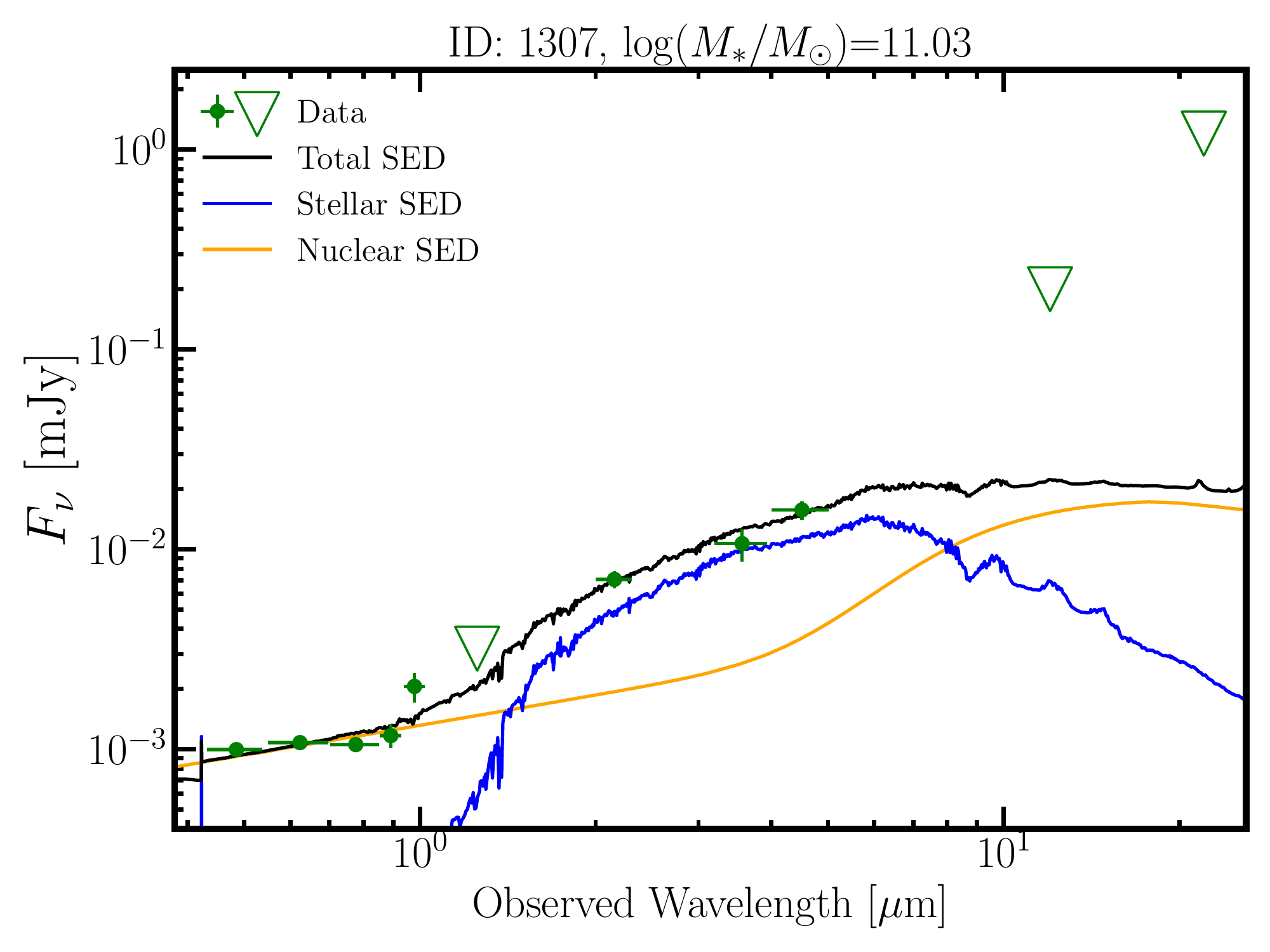}
\end{center}
\caption{Example of the SED fitting result on one of the HETDEX type 1 AGN. The photometric data used for fitting are shown with green circles with error bars. For the non-detections, we indicate the 1$\sigma$ upper limits with green open triangles. The best-fit composite SED is presented with the black solid curve, while the blue (orange) curve shows the stellar (nuclear) component. \label{fig:sed_eg}}
\end{figure}

For each HETDEX AGN, we fit the photometric data obtained in Section \ref{subsec:phot} and obtain the best-fit SED. For the non-detected photometric data points, we apply the $1\sigma$ upper limit flag in {\tt CIGALE}. Figure \ref{fig:sed_eg} shows an example of the best-fit SED. We confirm that 66/77 objects have  moderately well fitted SEDs with successfully de-composed AGN components and reduced $\chi^2<5$. We use these 66 AGN in our analysis. For the 11 excluded objects, one has a high reduced $\chi^2$ of 5.93, while the other 10 objects have best-fit AGN fraction of zero, indicating that either the objects are dominated by stellar light or the AGN component cannot be explained by the typical AGN model in Table \ref{tab:CIGALE}. 

\begin{figure*}[ht!]
\begin{center}
\includegraphics[scale=0.5]{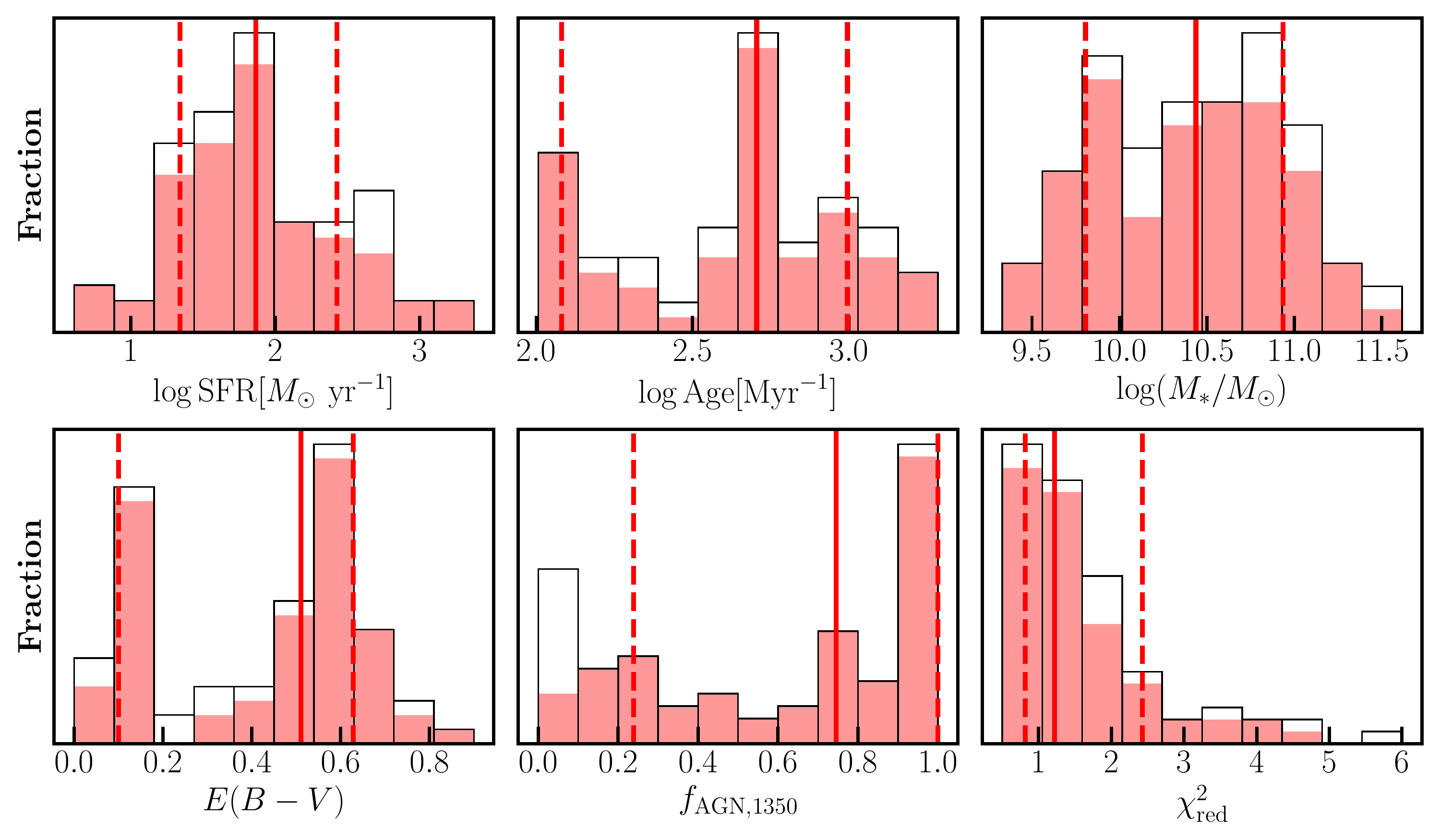}
\end{center}
\caption{Distributions of the best-fit SED parameters of the HETDEX type 1 AGN. Black open and red histograms indicate the distributions of all 77 objects and the 66 selected objects with good fitting results, respectively. Solid (dashed) vertical lines indicate the median values (16-84th percentiles) of the distributions for the 66 selected objects. \label{fig:sedparam_stellar}}
\end{figure*}

We examine the distributions of several key physical and fitting parameters as shown in Figure \ref{fig:sedparam_stellar}. The median values of the star formation rate (SFR), stellar age, $\mathcal{M_*}$, attenuation, and AGN fraction at rest-frame 1350\AA ($f_\mathrm{AGN,1350}$) are $\langle \mathrm{SFR}\rangle = 73.55_{-51.62}^{+194.39} M_\odot~\mathrm{yr}^{-1}$, $\langle \mathrm{Age}\rangle = 507.52_{-387.03}^{+482.04}$, $\langle \log(\mathcal{M_*}/M_\odot)\rangle = 10.44_{-0.63}^{+0.50}$, $\langle E(B-V)\rangle = 0.51_{-0.41}^{+0.11}$, $\langle f_\mathrm{AGN,1350}\rangle = 0.74_{-0.51}^{+0.25}$, where the errors refer to the 16th to 84th percentiles of the distributions. We find the distributions of SFR, stellar age, $\mathcal{M_*}$, and attenuation are consistent with those obtained in \citet{hainline12}, who studied the host galaxy properties of UV-selected AGN at $z\sim 2$. It should be noted that, the SFR derived from SED fitting with the absence of IR data may have a larger scatter of up to 0.3~dex \citep[e.g.,][]{ciesla15,florez20}. The typical fitting error in $\log\mathcal{M}_*$ is $0.3$~dex.

\begin{deluxetable*}{lcl}[htbp!]
\tablecaption{Parameters of the {\tt CIGALE} SED fitting\tablenotemark{a} \label{tab:CIGALE}}
\tablewidth{0pt}
\tablehead{
\colhead{Parameter} & 
\colhead{Value} &
\colhead{Description}
}
\startdata
\multicolumn{3}{c}{Star formation history} \\ 
age [Myr] & \textbf{100}, 158, \textbf{251}, 398, \textbf{631}, 1000, \textbf{1584}, 2512 & Age of the main stellar population in the galaxy \\ 
$\tau$ [Myr] & \textbf{100}, 158, \textbf{251}, 398, \textbf{631}, 1000, & e-folding time of the main stellar population in the galaxy \\
& \textbf{1584}, 2512, 3981, 6309, \textbf{10000} & \\
\hline
\multicolumn{3}{c}{Single stellar population \citep{bc03}} \\ 
IMF & \citet{chabrier03} & Initial mass function \\ 
metallicity & \textbf{0.02} & Metallicity (0.02 is solar) \\ \hline
\multicolumn{3}{c}{Dust attenuation \citep{calzetti00}} \\ 
$E(B-V)_\ast$ & 0.0, 0.1, \textbf{0.2}, 0.3, 0.4, 0.6, 0.8, 1.0 & Color excess of the stellar continuum light \\ \hline
\multicolumn{3}{c}{Dust emission \citep{dale14}} \\ 
$\alpha_\mathrm{SF}$ & 1.5, \textbf{2.0}, 2.5 & Power-law slope $\mathrm{d}U/\mathrm{d}M \propto U^{\alpha_\mathrm{SF}}$ \\ \hline
\multicolumn{3}{c}{Nebular \citep{inoue11}} \\ 
$\log U$ & \textbf{-2.0} & Ionisation parameter \\ \hline
\multicolumn{3}{c}{AGN emission \citep{fritz06}} \\ 
$R_\mathrm{max}/R_\mathrm{min}$ & \textbf{60} & Ratio of the maximum to minimum radii of the dust torus \\
$\tau_{9.7}$ & \textbf{6.0} & Optical depth at 9.7 microns \\
$\beta$ & \textbf{-0.5} & Radial dust distribution \\
$\gamma$ & \textbf{0.0} & Angular dust distribution \\
$\theta$ [deg.] & \textbf{100} & Full opening angle of the dust torus (Fig. 1 of \citealt{fritz06}) \\ 
$\psi$ [deg.] & \textbf{89.99} & Angle between equatorial axis and line of sight \\
$f_\mathrm{AGN}$ & 0.0, \textbf{0.05}, \textbf{0.1}, 0.15, 0.2, \textbf{0.25}, 0.3, 0.35, 0.4, 0.45, 0.5,  & Ratio of AGN IR luminosity to the total IR luminosity \\
 & 0.55, \textbf{0.6}, 0.65, 0.7, 0.75, 0.8, 0.85, 0.9, 0.95, 0.99 & \\
\enddata
\tablenotetext{a}{Numbers in bold font are used for the Monte Carlo simulations described in Section \ref{subsec:systematics}.}
\end{deluxetable*}

\subsection{$M_\mathrm{BH}$ measurements} \label{subsec:mbh}

\begin{figure}[ht!]
\begin{center}
\includegraphics[scale=0.38]{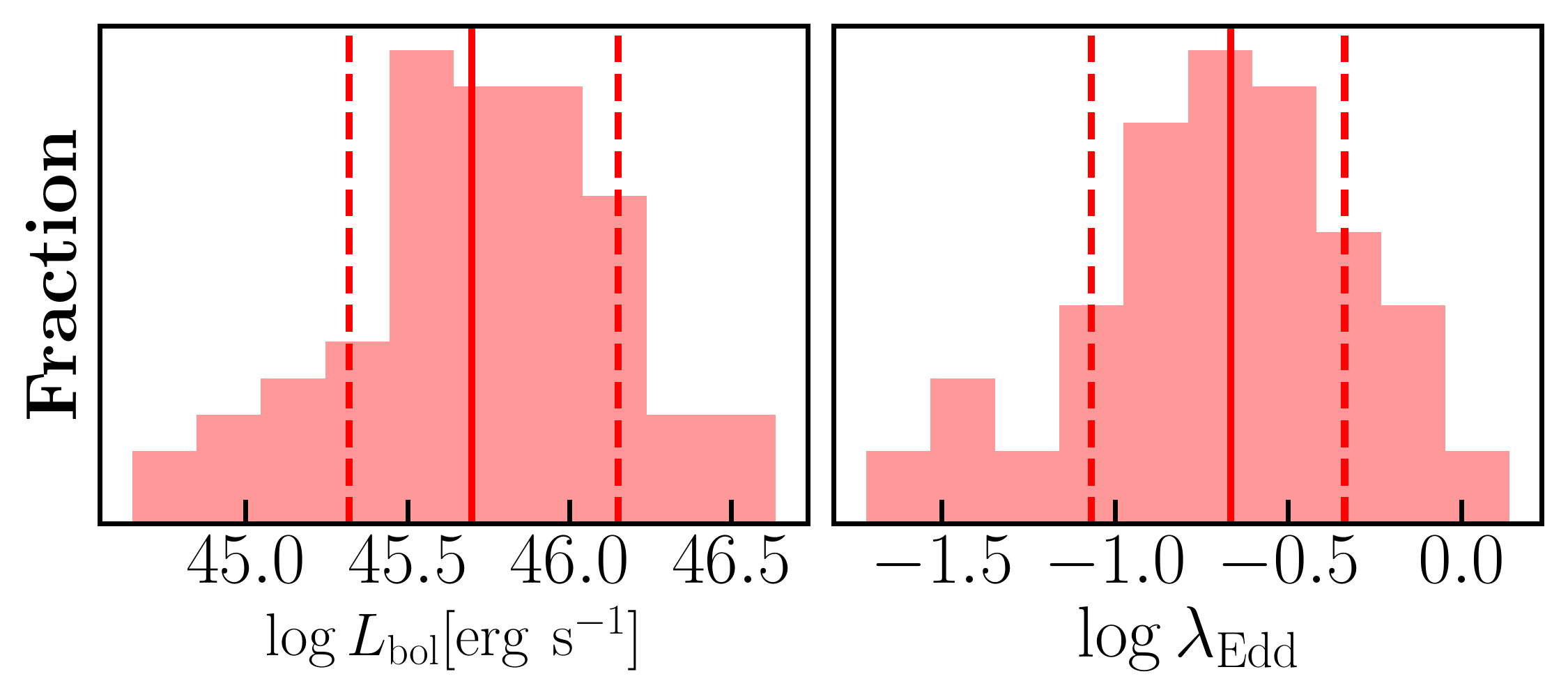}
\end{center}
\caption{Distributions of bolometric luminosities and Eddington ratios of our 66 selected type 1 AGN. Solid (dashed) vertical lines indicate the median values (16-84th percentiles) of the distributions. \label{fig:sedparam_agn}}
\end{figure}

We measure $M_\mathrm{BH}$ of HETDEX AGN with the single-epoch virial method using the C{\sc iv} estimator:
\begin{multline}
\label{eq:Mhb}
    \log\left(\frac{M_\mathrm{BH}}{M_\odot}\right) = A + B\log\left(\frac{L_\mathrm{1350}}{10^{44}\ \mathrm{erg\ s}^{-1}}\right) \\
    + 2\log\left(\frac{\mathrm{FWHM_{CIV}}}{\mathrm{1000~km\ s}^{-1}}\right),
\end{multline}
where FWHM$_\mathrm{CIV}$ and $L_{1350}$ are the C{\sc iv} line widths and monochromatic luminosities at rest-frame 1350~\AA, respectively. We adopt the parameters $(A,B)$ of $(6.66,0.53)$ from \citet{VP06}. We use the C{\sc iv} line widths from the HETDEX AGN catalog \citep{liu22b} and account for the instumental broadening \citep{hill21,gebhardt21}. For $L_{1350}$, we take the AGN components of the best-fit SED derived in Section \ref{subsec:mstar} and interpolate at the rest-frame 1350~\AA. This should remove the contamination of the host galaxy fluxes from the AGN fluxes. The virial $M_\mathrm{BH}$ derived with Eq. \ref{eq:Mhb} is known to have a scatter of $0.4$~dex \citep[e.g., ][]{shen12b,park17}, which is caused by non-virial motion, such as turbulence and outflows, in the C{\sc iv}-emitting broad line region \citep[e.g.,][]{kz11,kz13}. Figure \ref{fig:sedparam_agn} shows the distributions of the derived bolometric luminosities ($L_\mathrm{bol}$) and Eddington ratios ($\lambda$) of our AGN sample. The median values of $L_\mathrm{bol}$ and $\lambda$ are $\langle\log (L_\mathrm{bol}/[\mathrm{erg~s}^{-1}])\rangle = 45.69_{-0.38}^{+0.45}$, $\langle\log\lambda\rangle = -0.67_{-0.40}^{+0.33}$, where the errors are defined as the 16th to 84th percentile of the distributions. Our $\langle\lambda\rangle$ agrees with the results in \citet{aird18}, who found the median specific BH accretion rate ($\lambda_\mathrm{sBHAR}$), which is defined to be equivalent to $\lambda$, of X-ray detected AGN at $z=2.0-2.5$ to be $\log\lambda_\mathrm{sBHAR}\sim-0.5$ at the $\log(\mathcal{M_*}/M_\odot)$ range of 10.0-10.5.

\subsection{Systematics}
\label{subsec:systematics}
\begin{figure*}[ht!]
\begin{center}
\includegraphics[scale=0.47]{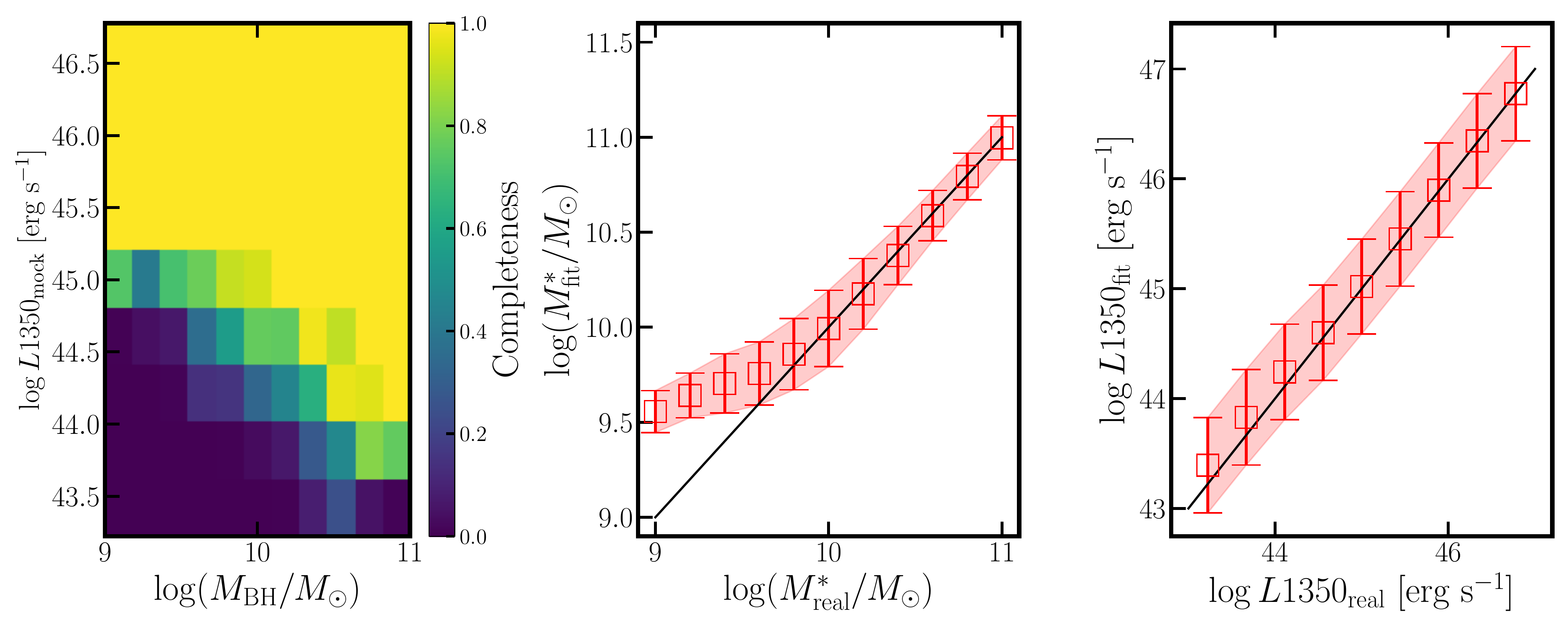}
\end{center}
\caption{Monte Carlo Simulations on the SEDs of mock objects created with CIGALE. Left: The $K_s$-band 5$\sigma$ detection fraction as a function of input stellar masses and input AGN UV luminosities simulated with the ``saveflux" method in \texttt{CIGALE}. Middle: Comparison between the input stellar mass ($\log\mathcal{M}_\mathrm{real}^*$) and the best-fit stellar mass ($\log\mathcal{M}_\mathrm{fit}^*$). The red open squares and error bars indicate the median and 1$\sigma$ scatter, respectively. The black solid line shows the one-to-one relation. Right: Same as the middle panel, but for the AGN UV luminosity.\label{fig:cig_sim}}
\end{figure*}

Here we discuss the potential systematics in the $\mathcal{M}_*$ and $\mathcal{M}_\mathrm{BH}$ estimations. We conduct Monte Carlo simulations
to examine the performance of host-nuclear decomposition with \texttt{CIGALE}. We first generate mock SEDs and their respective mock observational fluxes in each of the filters (Table \ref{tab:multiband}) using the "savefluxes" method provided by \texttt{CIGALE}. We normalize the stellar SEDs assuming stellar masses of $\log \mathcal{M}_*/\mathcal{M}_\odot=9.0-11.0$ with a 0.4~dex interval. At each $\log \mathcal{M}_*$, we generate mock SEDs with the parameters indicated in bold in Table \ref{tab:CIGALE}, yielding $4\times 5\times 4=80$ different shapes of SEDs at a given combination of ($\log \mathcal{M}_*$, $f_\mathrm{AGN}$). For each mock SED, we then add the actual $1\sigma$ observational error to the generated mock observed fluxes and make $500$ mock catalogs with a Gaussian probability distribution that has a standard deviation of the $1\sigma$ observational errors. If the mock observation in a specific band has a flux density fainter than the limiting magnitude, we put a $1\sigma$ upper limit to mimic the treatment of non-detections in our data catalogs. After making the mock observational catalog of $80\times 500 = 40000$ objects, we conduct the SED fitting on these objects in the same manner as mentioned in Section \ref{subsec:mstar}. The comparisons between the input $\mathcal{M}_*$($L_{1350}$) to the output $\mathcal{M}_*$($L_{1350}$) are shown in Figure \ref{fig:cig_sim}. We find that our method can successfully reproduce the $L_{1350}$ with a scatter of 0.4~dex. The $\mathcal{M}_*$ of our objects are also well reproduced within a scatter of 0.2~dex when the input $\log \mathcal{M}_*/\mathcal{M}_\odot$ is greater than 9.6. At $\log \mathcal{M}_*/\mathcal{M}_\odot < 9.6$, the SED fitting starts to overestimate the $\mathcal{M}_*$ mainly due to the depth of our photometric data especially in the NIR bands. These measurement errors and systematics will be included in our forward modelling analysis in Section \ref{sec:model}.

\subsection{Comparison with JWST image decomposition}\label{subsec:jwst}
\begin{figure*}[ht!]
\begin{center}
\includegraphics[scale=0.22]{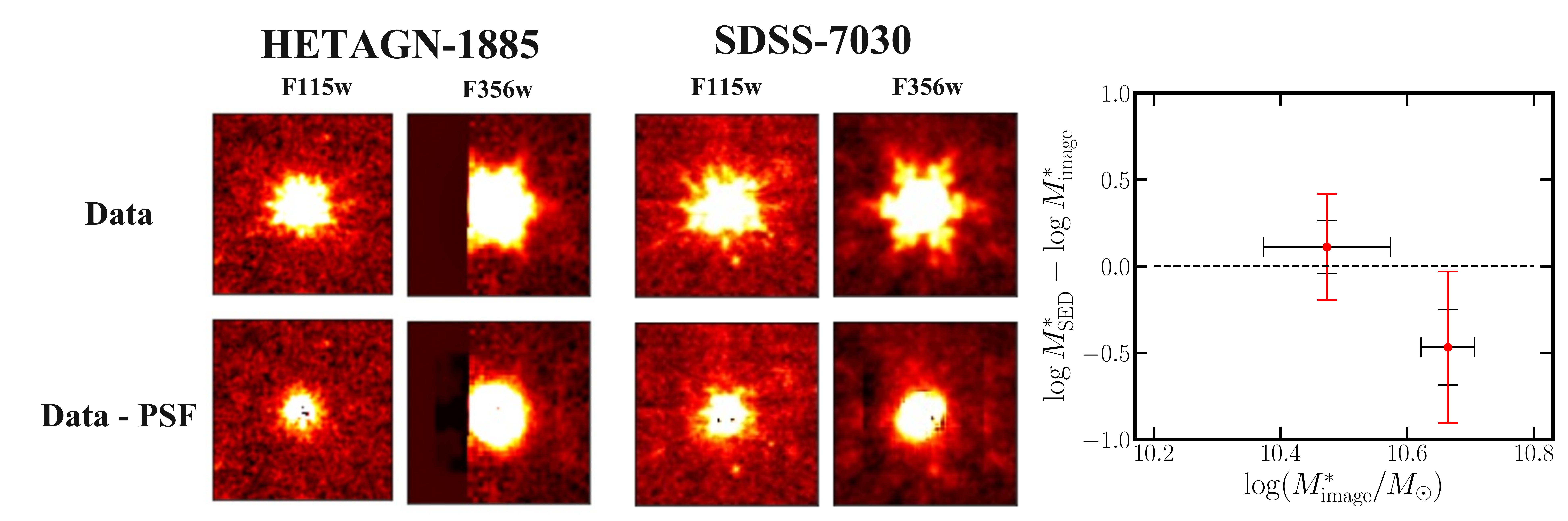}
\end{center}
\caption{Left and Middle: Image decomposition of two type~1 AGN with JWST CEERS data. From top to bottom are the observed data and data subtracting the modeled PSF (i.e. host galaxy without nuclear flux). Right: Comparison between the $\mathcal{M}_*$ derived with image and SED decomposition. The black and red error bars indicate the 1$\sigma$ and 2$\sigma$ errors, respectively. \label{fig:galfit}}
\end{figure*}

At $z<2$, the host-nuclear decompositions and $\mathcal{M}_*$ estimations of type~1 AGN have been performed through two-dimensional surface brightness profile modelling on the resolved host galaxy images taken with ground-based telescopes and Hubble Space Telescope (HST). Proir to the JWST era, the same analyses cannot be conducted for $z>2$ AGN due to the lack of imaging data that can spatially resolve the host galaxies at the rest-frame optical wavelengths. Utilizing the latest JWST/NIRCam ERS data released in June 2022, here we assess the reliability of our SED decomposition by comparing our results with those derived from the image decomposition based on JWST/NIRCam imaging. 

We use the data taken from the Cosmic Evolution Early Release Science \citep[CEERS;][]{finkelstein17,finkelstein22,bagley22} that was released in June 2022. CEERS data include four pointings covering 33.1~arcmin$^2$, with seven bands of $F115W$, $F150W$, $F200W$, $F277W$, $F356W$, $F410M$, and $F444W$. Details of the image reduction are described in \citet{harikane22}. In the reduced CEERS imaging we find one type 1 AGN, HETAGN-1885, from the HETDEX AGN catalog, and one type 1 AGN, SDSS-7030, from the SDSS DR14 QSO catalog \citep{rakshit20} using the same selection criteria mentioned in Section \ref{sec:data}. 

We conduct the two-dimensional host-nuclear decomposition by fitting a Sersic profile and a point spread function (PSF) to the JWST images of the two AGN in all seven bands. 
For each band, we generate the PSF by selecting and stacking bright stars in the same fields. We then fit a Sersic profile and a PSF simultaneously to the object images. For the Sersic profile, we restrict the Sersic index $n$ to the range of $1-4$. We conduct the fitting in all of the seven filters, and select the filter with the best reduced-$\chi^2$ that is located at a wavelength redder than the rest-frame 5000~\AA where the stellar components are the most prominent. Assuming the stellar population is the same, we then fix the effective radius ($r_e$), axis ratio, $n$, and position angle to the best-fit results in that band and conduct the fitting again to all the other filters. The results are shown in Figure \ref{fig:galfit}. For comparison, we also fit a single PSF to the objects in Figure \ref{fig:galfit}. 
We find that for both AGN, the additional Sersic profile is necessary in all seven filters. We then derive the $\mathcal{M}_*$ by conducting SED fittings with the flux densities of stellar component indicated in Table \ref{tab:CIGALE} using \texttt{CIGALE}. We use only the components of stellar continua, nebular emissions, dust attenuation, and dust emission models during the fitting, while the input parameters are the same as Table \ref{tab:CIGALE}. The resulting $\mathcal{M}_*$ for HETAGN-1885 and SDSS-7030 are $\log(\mathcal{M}_*/M_\odot) = 10.47\pm0.10$ and $10.66\pm0.04$, respectively. Since we do not assume the AGN component for the SED fitting, we estimate the upper limit of $L_{1350}$ for the two objects based on the multiband catalog of \citet{stefano17} and derive the $\mathcal{M}_\mathrm{BH}$ for HETAGN-1885 and SDSS-7030 to be $\log(\mathcal{M}_\mathrm{BH}/M_\odot) < 8.38$ and $8.89$, respectively (Figure \ref{fig:mago_obs}).

We compare the obtained $\mathcal{M}_*$ with those derived from the SED decomposition, and show the results in Figure \ref{fig:galfit}. 
For HETAGN-1885, $\mathcal{M}_*$ derived with JWST image decomposition and SED decomposition are consistent within 0.2~dex. For SDSS-7030, JWST image decomposition yield to a 0.5~dex higher $\mathcal{M}_*$ than the SED decomposition, which corresponds to a $2\sigma$ consistency. 
Despite the small sample size of only two objects, our results based on the availeble data suggest that the host-nuclear decomposition with \texttt{CIGALE} is marginally consistent with those given by the image decomposition.

\section{Observed $\mathcal{M_*}-\mathcal{M}_\mathrm{BH}$ Relation} \label{sec:obs_results}
\begin{figure}[h!]
\begin{center}
\includegraphics[scale=0.47]{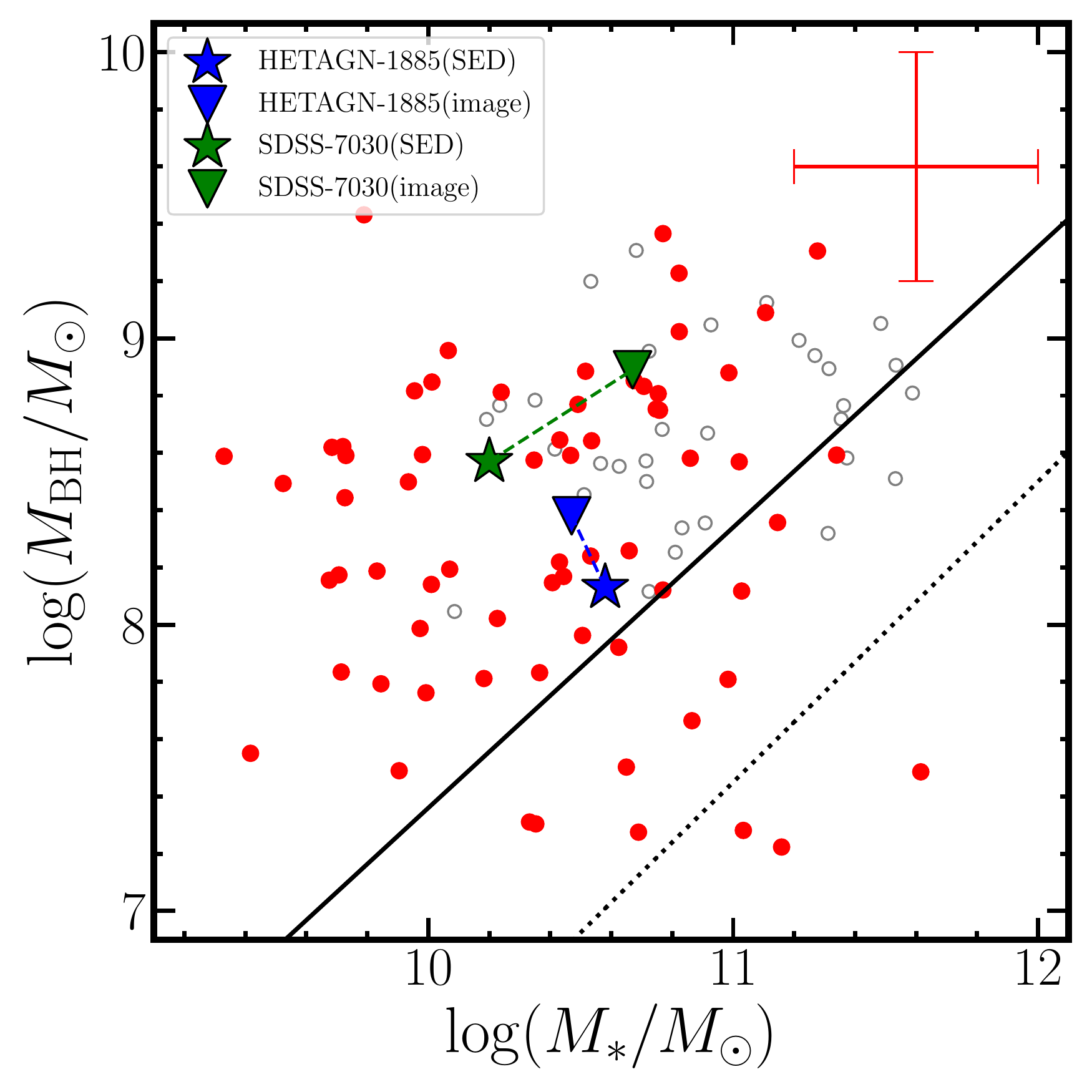}
\end{center}
\caption{Observed $\mathcal{M}_*-\mathcal{M}_\mathrm{BH}$ relation for HETDEX type 1 AGN (red circles). The typical errors in $\mathcal{M}_*$ and $\mathcal{M}_\mathrm{BH}$ are shown at the bottom-left. The two AGN with JWST CEERS imaging data (Section \ref{subsec:jwst}), HETAGN-1885 and SDSS-7030, are denoted in blue and green symbols, respectively. We use Stars (triangles) to represent the $\mathcal{M}_*$ and $\mathcal{M}_\mathrm{BH}$ obtained through imaging decomposition (\texttt{CIGALE} SED decomposition). Previous observational results at $z\sim 2$ from \citet{merloni10} and \citet{sun15} compiled by \citet{yang18} are indicated with the open grey circles. The black solid line shows the local $\mathcal{M}_*-\mathcal{M}_\mathrm{BH}$ relation of both AGN and relic BHs fitted by \citet{ding20}. The local relation obtained by \citet{rv15} based on AGN only is denoted with black dotted line. \label{fig:mago_obs}}
\end{figure}

We present the observed $\mathcal{M}_*-\mathcal{M}_\mathrm{BH}$ relation for our HETDEX type 1 AGN in Figure \ref{fig:mago_obs}, together with two additional objects mentioned in Section \ref{subsec:jwst} and previous results at the similar redshift range compiled by \citet{yang18}. The typical error in $\mathcal{M}_*$ ($\mathcal{M}_\mathrm{BH}$) shown at the bottom-left of the figure are estimated with the quadrature sum of the typical measurement error mentioned in Section \ref{subsec:mstar} (\ref{subsec:mbh}) and the systematic scatter derived in Section \ref{subsec:systematics}. Our HETDEX type 1 AGN cover the mass range of $9.4-11.6$ and $7.2-9.8$ in $\log (\mathcal{M}_*/\mathcal{M}_\odot)$ and $\log (\mathcal{M}_\mathrm{BH}/\mathcal{M}_\odot)$, with the median of 10.44 and 8.27, respectively. Compared with previous results at the same redshift range from \citet{merloni10,sun15} compiled by \citet{yang18}, the mass ranges probed in our study are $\sim 0.8$~dex smaller in $\mathcal{M}_\mathrm{BH}$ and $\sim 0.6$~dex smaller in $\mathcal{M}_*$. 

In Figure \ref{fig:mago_obs} we also compare our sample at $z\sim2$ with the local relation of \citet{ding20} and \citet{rv15} with the functional form of 
\begin{equation} \label{eq:mago}
    \log\mathcal{M}_\mathrm{BH} = c_1 \log\mathcal{M}_* + c_2,
\end{equation}
as plotted in the black solid and dotted line, respectively. The local relation of \citet{ding20}, which has $c_1=0.98$ and $c_2=-2.56$, is derived by compiling and fitting the local relic BH sample of \citet{hr04} and AGN sample of \citet{bennert10,bennert11} with $6.5 < \log(\mathcal{M}_\mathrm{BH}/M_\odot) < 9.5$, and thus represents the local relation of both active an inactive BHs. The \citet{rv15} relation, which is $\sim1$~dex lower than the \citet{ding20} relation, is based on type~1 AGN with $\mathcal{M}_\mathrm{BH}$ range of $5 < \log(\mathcal{M}_\mathrm{BH}/M_\odot) < 8$. Similar to the previous results at $z\sim2$ compiled by \citet{yang18}, the majority of our HETDEX type 1 AGN are located above the local relations except for a few outliers. Taking into account the errors in mass measurements of our sample as well as the intrinsic scatter of the local relations, we find that 86\% (57/66) of our type 1 AGN lie above the \citet{rv15} relation. Even if comparing with the \citet{ding20} relation with higher normalization, more than half of our sample still locate above the local anchor, indicating that our sample is likely to represent AGN with overmassive black holes. Because the \citet{ding20} is derived with the local BHs that has a similar $\mathcal{M}_\mathrm{BH}$ range to ours, we use the \citet{ding20} relation as the local anchor for the following analysis.

We fit Equation (\ref{eq:mago}) to the observed $\mathcal{M}_* - \mathcal{M}_\mathrm{BH}$ relation by maximizing the log-likelihood function that account for errors in both $\log\mathcal{M}_*$ ($\sigma_*$) and $\log\mathcal{M}_\mathrm{BH}$($\sigma_\mathrm{BH}$): 
\begin{multline} 
\label{eq:loglikelihood}
    \log\mathcal{L} = \sum\Bigg[\frac{(\log\mathcal{M}_\mathrm{BH} - c_1\log\mathcal{M}_*-c_2)^2}{\sigma_\mathrm{BH}^2+c_1^2\sigma_*^2} \\ + \ln\left(\sigma_\mathrm{BH}^2+c_1^2\sigma_*^2\right)\Bigg].
\end{multline}
Given the large uncertainties in both $\mathcal{M}_*$ and $\mathcal{M}_\mathrm{BH}$ measurements, as well as the intrinsic scatter of $\mathcal{M}_* - \mathcal{M}_\mathrm{BH}$ relation, we do not find 
any tight correlation with a correlation coefficient of $0.05$. 
Such a non-correlation for the observed $\mathcal{M}_*$ and $\mathcal{M}_\mathrm{BH}$ was also found in previous results at high-z \citep[e.g.,][]{merloni10}, as the relatively large measurement uncertainties compared with low-z are likely to wash away the intrinsic correlation \citep{kelly07}. Following the approaches of previous studies, we assume a linear relation with a fixed slope identical to the local relation \citep{ding20}, i.e., $c_1=0.98$. The best-fit $c_2$ value is $-1.86\pm0.07$, indicating that our observed $\mathcal{M}_* - \mathcal{M}_\mathrm{BH}$ relation has a 0.58~dex positive offset from the local relation. The positive offset, however, may or may not be the consequence of selection biases, which will be addressed in the following section.

\section{Forward Modelling and Intrinsic $\mathcal{M}_*-\mathcal{M}_\mathrm{BH}$ Relation} \label{sec:model}
As mentioned in Section \ref{sec:obs_results}, the observed $\mathcal{M}_*-\mathcal{M}_\mathrm{BH}$ relations would have various observational and selection biases that may result in the offset from the local relation. To account for such biases, we apply a Monte Carlo simulation that is similar to \citet{sun15} and \citet{li21b} to infer the intrinsic scaling relation at $z\sim2$.
In general, for a given intrinsic relation, we generate a mock AGN catalog that incorporates the actual observational and selection effects. Using the distribution of the mock AGN on the $\mathcal{M}_*-\mathcal{M}_\mathrm{BH}$ plane as the probability distribution, we calculate the sum of log-likelihood of our observed type 1 AGN. By maximizing the log-likelihood we find the parameters of the best-fit intrinsic relation.

Specifically, we first assign $\mathcal{M}_*$ values to the mock AGN, randomly sampling the stellar mass function (SMF) of \citet{davidzon17} that is in the form of double Schechter function:
\begin{multline} 
\label{eq:dSch}
    \Phi (\mathcal{M}_*) \, d\mathcal{M_*} = \left[\Phi^*_1 \left(\frac{\mathcal{M}_*}{\mathcal{M}_{*,c}}\right)^{\alpha_1} + \Phi^*_2 \left(\frac{\mathcal{M_*}}{\mathcal{M}_{*,c}}\right)^{\alpha_2} \right] \\
    \exp\left(-\frac{\mathcal{M}_*}{\mathcal{M}_{*,c}}\right) \frac{d\mathcal{M}_*}{\mathcal{M}_*},
\end{multline}
where we use the parameters $\mathcal{M}_{*,c}, \alpha_1, \Phi^*_1, \alpha_2, \Phi^*_2$ at $z=2.0-2.5$ listed in Table 1 of \citet{davidzon17}. The stellar masses generated in this step is referred to as $\mathcal{M}_\mathrm{*,true}$.
Next, we assign the $\mathcal{M}_\mathrm{BH}$ to the mock AGN based on $\mathcal{M}_\mathrm{*,true}$, the input intrinsic $\mathcal{M}_*-\mathcal{M}_\mathrm{BH}$ relation, and the input intrinsic scatter $\sigma_\mu$ that follows a Gaussian distribution. The probability distribution of $\mathcal{M}_\mathrm{BH}$ is thus given by:
\begin{multline} \label{eq:ms_mbh}
    P(\log\mathcal{M}_\mathrm{BH}) = \frac{1}{\sqrt{2\pi}\sigma_\mu}\\
    \exp \left(-\frac{(\log\mathcal{M}_\mathrm{BH}-(c_1\log\mathcal{M}_*+c_2))^2}{2\sigma_\mu^2}\right).
\end{multline}
We refer to the BH masses assigned in this step as $\mathcal{M}_\mathrm{BH,true}$. The galaxies generated in the above steps include both active galaxies observed as AGN and inactive galaxies with relic BHs. Here we assume that the underlying $\mathcal{M}_*$ and $\mathcal{M}_\mathrm{BH}$ distributions of AGN are the same as galaxies at $z\sim2$. The potential impact of such an assumption will be further discussed in Section \ref{subsec:assumptions}.
Because the AGN selection biases are mainly due to the observational cuts on the luminosities, we also assign $\lambda$ and hence bolometric luminosities ($L_\mathrm{bol,true}$) to the mock AGN by assuming the intrinsic Eddington ratio distribution function (ERDF). We apply the ERDF of \citet{schulze15} that is in the form of a Schechter function \citep{schechter76}:
\begin{equation} \label{eq:Sch}
    \phi (\lambda) = \frac{\phi^*}{\lambda_*} \left(\frac{\lambda}{\lambda_*}\right)^{\alpha_\lambda}\exp\left(-\frac{\lambda}{\lambda_*}\right),
\end{equation}
For the redshift-dependant parameters ($\phi^*, \lambda_*, \alpha_\lambda$), we use the values given in the Table 1 of \citet{schulze15}. Because the ERDF of \citet{schulze15} cover the $\mathcal{M}_\mathrm{BH}$ range of $8.0<\log(\mathcal{M}_\mathrm{BH}/M_\odot)<10.0$, we extrapolate the ERDF with the assumption that SMBHs with lower $\mathcal{M}_\mathrm{BH}$ follow the same Eddington ratio distribution. With $\mathcal{M}_\mathrm{BH,true}$ and $\lambda$ we calculate $L_\mathrm{bol,true}$ for each object, and convert $L_\mathrm{bol,true}$ to the intrinsic monochromatic luminosity at rest-frame 1350~\AA\ ($L_\mathrm{1350, true}$) using the bolometric correction factor $L_\mathrm{bol} = 3.81\times L_{1350}$ from \citet{richards06a}. We insert $\mathcal{M}_\mathrm{BH, true}$ and $L_\mathrm{1350, true}$ into Equation \ref{eq:Mhb} to obtain the intrinsic C{\sc iv} line width, FWHM$_\mathrm{CIV, true}$. We also assign the intrinsic C{\sc iv} emission line flux ($F_\mathrm{CIV, true}$) to each mock AGN from the C{\sc iv} equivalent width (EW) distribution and $L_\mathrm{1350, true}$. We use the C{\sc iv} EW distribution of SDSS DR14 QSOs at $z=2-2.5$ that follows the log-normal distribution \citep{rakshit20}.

After obtaining the intrinsic physical properties and observables ($\mathcal{M}_\mathrm{*,true}, \mathcal{M}_\mathrm{BH,true}, \lambda, L_\mathrm{1350, true}$, FWHM$_\mathrm{CIV, true}$, $F_\mathrm{CIV, true}$) of the mock AGN catalog, we add the measurement biases. For the measurement bias in $\mathcal{M_*}$ ($L_{1350}$), we add scatter to each sampled $\mathcal{M}_\mathrm{*,true}$ ($L_\mathrm{1350, true}$) by randomly assigning a $\mathcal{M_*}$ ($L_{1350}$) that is drawn from a Gaussian distribution with the mean and standard deviation indicated in Figure \ref{fig:cig_sim}, obtaining the observed stellar mass (UV continuum luminosity), $\mathcal{M}_\mathrm{*,obs}$ ($L_\mathrm{1350, obs}$).
For the measurement bias in C{\sc iv} virial mass estimation, we randomly generate the observed C{\sc iv} line width (FWHM$_\mathrm{CIV, obs}$) by adding the scatter to each FWHM$_\mathrm{CIV, true}$ with a standard deviation of 0.4~dex \citep[e.g., ][]{shen12b,park17}. We then insert $L_\mathrm{1350, obs}$ and FWHM$_\mathrm{CIV, obs}$ back into Equation \ref{eq:Mhb} to obtain the``observed" BH mass $m_\mathrm{obs}$.

Finally, with the ``observed" physical properties of the mock AGN catalog, we apply the observational cuts on continuua and C{\sc iv} emission lines with the same selection functions as applied to our HETDEX type 1 AGN. Namely, from $L_\mathrm{1350, obs}$ and $\mathcal{M}_\mathrm{*,obs}$ we randomly select objects according to the selection function given in Figure \ref{fig:cig_sim} that is simulated based on $K_\mathrm{s}$-band detection criteria. For the C{\sc iv} emission lines, we require FWHM$_\mathrm{CIV, obs} > 1000$~km~s$^{-1}$ and $F_\mathrm{CIV, obs}$ to be greater than $5\sigma$ of the HETDEX spectral noise. 

\subsection{Intrinsic $\mathcal{M_*}-\mathcal{M}_\mathrm{BH}$ relation} \label{sec:int_results}
\begin{figure*}[ht!]
\begin{center}
\includegraphics[scale=0.8]{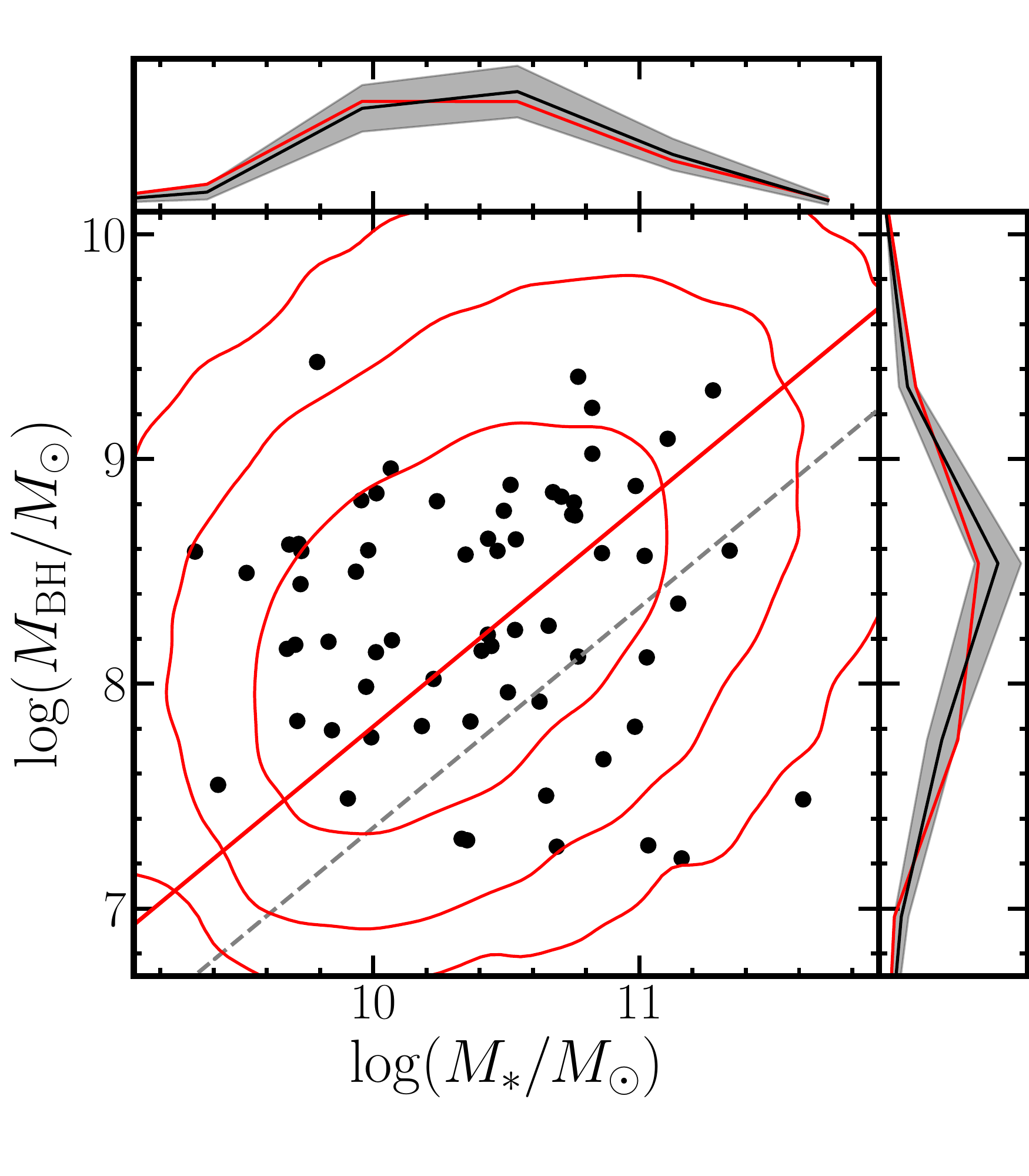}
\end{center}
\caption{Best-fit intrinsic  $\mathcal{M}_*-\mathcal{M}_\mathrm{BH}$ relation (red solid line) to the observed HETDEX type 1 AGN (black circles), with slope fixed to 0.98 to match \citet{ding20}. The red solid contours demonstrate the $1, 2, 3\sigma$ distributions of the mock AGN sample (Section \ref{sec:model}) that is generated with the best-fit intrinsic relation with observational biases included. For comparison, we also show the local $\mathcal{M}_*-\mathcal{M}_\mathrm{BH}$ relation fitted by \citet{ding20} with the grey dashed line. In the top (right) panel, we also show the one-dimensional distribution in $\mathcal{M}_*$ ($\mathcal{M}_\mathrm{BH}$) of the observed HETDEX type 1 AGN (black solid curve with grey shaded area showing the 1$\sigma$ scatter) and the mock AGN sample (red curve). \label{fig:result}}
\end{figure*}

We apply our forward modelling to the observed $\mathcal{M_*}-\mathcal{M}_\mathrm{BH}$ relation of our HETDEX type 1 AGN sample (Figure \ref{fig:mago_obs}) and constrain the intrinsic relation. Due to the uncertainties in the C{\sc iv} $\mathcal{M}_\mathrm{BH}$ estimator, we fix the intrinsic scatter $\sigma_\mu$ to 0.3~dex and $c_1$ to 0.98 \citep{ding20}, both equivalent to the values for the local $\mathcal{M_*}-\mathcal{M}_\mathrm{BH}$ relation. We obtain the best-fit intrinsic $\mathcal{M_*}-\mathcal{M}_\mathrm{BH}$ relation with $c_2=-1.92\pm0.14$, which has a moderately positive offset of $0.52\pm0.14$~dex from the local relation (Figure \ref{fig:result}). Assuming that the $\mathcal{M_*}-\mathcal{M}_\mathrm{BH}$ relation evolves with the redshift in the form of $\Delta\log\mathcal{M}_\mathrm{BH} = \gamma\log(1+z)$, our result at $z=2.2$ would yield to a positive evolution of $\gamma=1.03\pm0.28$. Such a result is larger than the value of $\gamma=0.12_{-0.27}^{+0.28}$ at $z<0.8$ given by \citet{li21b} and $\gamma=0.55\pm0.15$ at $z=1.2-1.7$ given by \citet{ding20}. Compared with previous results at the similar redshift of $z\sim2$, our $\gamma$ is consistent with \citet{merloni10} at the massive end within 1$\sigma$ level, while extending both $\mathcal{M_*}$ and $\mathcal{M}_\mathrm{BH}$ to the lower-mass regime. 


\section{Discussion} \label{sec:discussion}
\subsection{Redshift Evolution of $\mathcal{M_*}-\mathcal{M}_\mathrm{BH}$ relation} \label{subsec:discussion}
\begin{figure}[ht!]
\begin{center}
\includegraphics[scale=0.45]{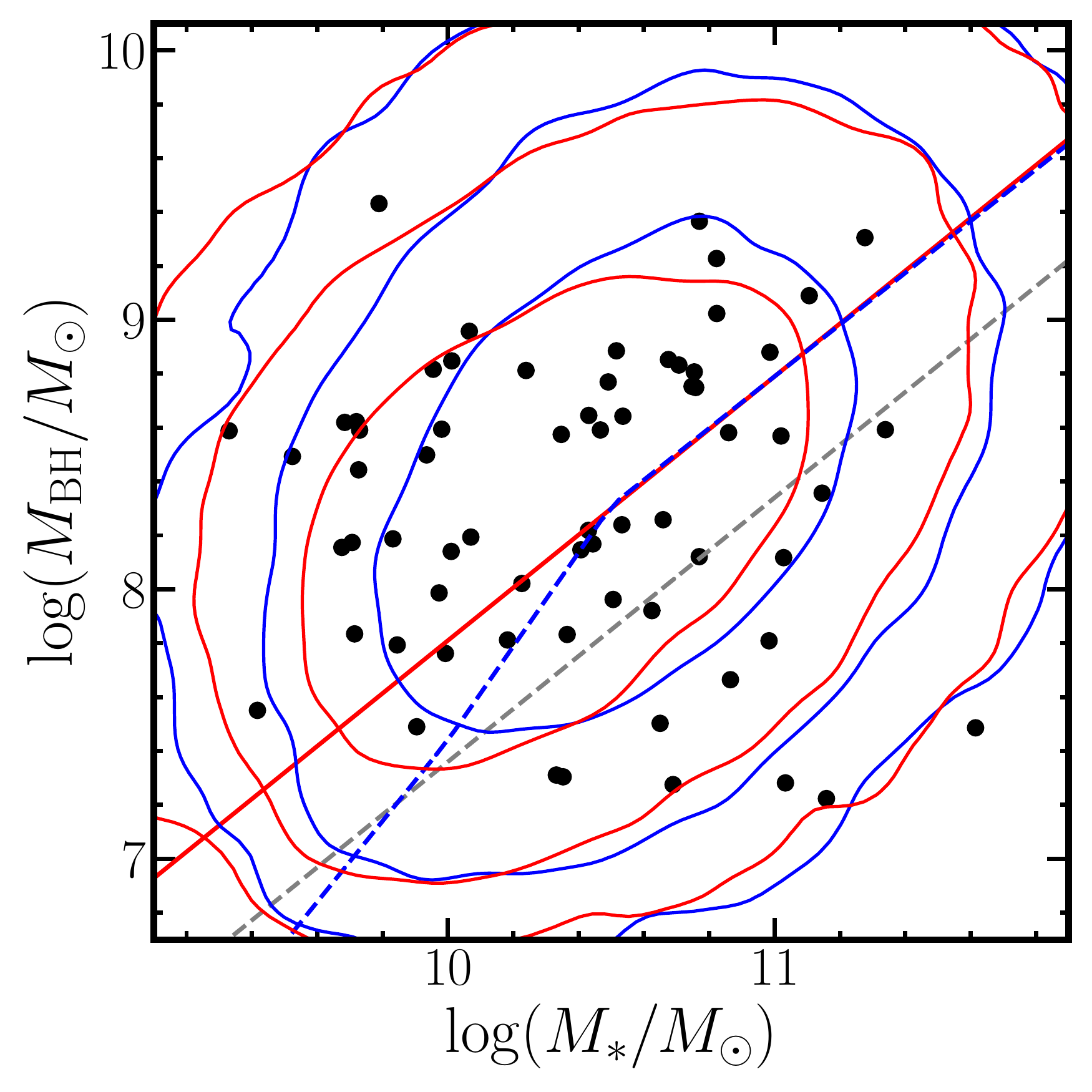}
\end{center}
\caption{Same as Figure \ref{fig:result}, but with the addition of the non-linear relation predicted by TNG100 simulation (blue dashed curve). The blue solid contours represent the $1, 2, 3\sigma$ intervals of the mock AGN sample (Section \ref{sec:model}) that is generated with the non-linear relation predicted by TNG100 model. \label{fig:result_comp}}
\end{figure}
We compare the redshift evolution of the intrinsic $\mathcal{M_*}-\mathcal{M}_\mathrm{BH}$ relation derived from our $z\sim 2$ observational results with those predicted by simulations, such as the hydrodynamic simulations of Illustris \citep{sijacki15}, IllustrisTNG \citep[hereafter TNG;][]{weinberger17,weinberger18}, SIMBA \citep{dave19,thomas19}, and Horizon \citep{volonteri16}, as well as the empirical model of Trinity \citep{zhangh21}. 
The comparison between different hydrodynamic models has been summarized in \citet{habouzit21}. For example, the Illustris, Horizon-AGN, and EAGLE simulations predict that the normalization of $\mathcal{M_*}-\mathcal{M}_\mathrm{BH}$ relation increases towards higher redshifts, suggesting that BH growth is more efficient at high redshifts, which is consistent with our results. Theoretically, galaxies at higher redshifts that are more compact and gas-rich can provide more fuel to the central SMBHs \citep[e.g.,][]{wellons15,habouzit19}, which is also supported by the observed tight correlation between the BH accretion rate and the compactness of host galaxies \citep[e.g.,][]{kocevski17,ni21}. Another possible explanation to such an increasing normalization with redshift is the transformation of the dominant contribution to $\mathcal{M}_*$ from disk to bulge \citep{ding20}. 
On the other hand, the TNG and SIMBA simulations predict that the normalization of $\mathcal{M_*}-\mathcal{M}_\mathrm{BH}$ relation becomes lower towards higher redshifts, due to BH growth being more efficient at lower redshifts.
A similar result is also given by \citet{zhangh21}, who applied an empirical model by fitting the observations. They found that SMBHs at higher redshifts tend to have $\mathcal{M}_\mathrm{BH}$ that fall below the local $\mathcal{M_*}-\mathcal{M}_\mathrm{BH}$ relation especially at the low-mass end, although the difference is small at $z<2$.

We also notice that for the TNG and EAGLE simulations, there are two phases of BH growth characterized by a non-linear $\mathcal{M_*}-\mathcal{M}_\mathrm{BH}$ relation that features a break at a critical stellar mass $\mathcal{M}_\mathrm{*,crit}\sim 10^{10.5}~M_\odot$. While the slope and normalization of the $\mathcal{M_*}-\mathcal{M}_\mathrm{BH}$ is similar to our best-fit linear relation (Section \ref{sec:int_results}), below
$\mathcal{M}_\mathrm{*,crit}$ the slope of $\mathcal{M_*}-\mathcal{M}_\mathrm{BH}$ relation becomes steeper than the slope beyond $\mathcal{M}_\mathrm{*,crit}$, indicating that BH growth is more suppressed in lower-mass systems. Such a suppression of $\mathcal{M}_\mathrm{BH}$ in the low-mass regime is caused by strong supernova (SN) feedback in low-mass galaxies that depletes the center of galaxies of cold gas and stunts early BH growth \citep[e.g.,][]{dubois15,aa17,bower17}. We test whether such a suppression of $\mathcal{M}_\mathrm{BH}$ at the low mass end can be observed with our sample by comparing the $\mathcal{M_*}-\mathcal{M}_\mathrm{BH}$ distribution of the mock AGN sample predicted by out best-fit linear relation in Section \ref{sec:int_results} and the TNG100 non-linear relation \citep[Table C2 in][]{habouzit21}, respectively, to that of our observed HETDEX type 1 AGN. Here we use the $\mathcal{M_*}-\mathcal{M}_\mathrm{BH}$ relation predicted by TNG100 because its slope at the massive end is almost the same as our best-fit linear relation. We also fix the normalization and intrinsic scatter to be identical to our best-fit linear result (Section \ref{sec:int_results}), so that the difference in the $\mathcal{M_*}-\mathcal{M}_\mathrm{BH}$ distributions, if any, would be caused by the break at $\mathcal{M}_\mathrm{*,crit}$. 
We then apply the forward modelling method in Section {\ref{sec:model}} to generate the mock AGN samples as shown in Figure \ref{fig:result_comp}. We find that even after accounting for the observational biases, the non-linear relation predicted by TNG100 simulation would yield somewhat fewer objects at the low-$\mathcal{M_*}$ end in the observed $\mathcal{M_*}-\mathcal{M}_\mathrm{BH}$ plane. We compare the goodness of fit given by these two models in Figure \ref{fig:result_comp} using the two-dimensional (2D) two-sample Kolmogorov-Smirnov (KS) test. The $p$-value of the 2D two-sample KS test between the best-fit linear (TNG100 non-linear) relation and the observed data is 0.10 (0.02), ruling out the hypothesis that our observed sample is drawn from the best-fit linear (TNG100 non-linear) relation at the 90\% (98\%) confidence level. Hence our results disfavor the TNG non-linear relation, suggesting less suppression of $\mathcal{M}_\mathrm{BH}$ i.e., weaker SN feedback at the low-mass end than occurs in the TNG100 simulation.

Interestingly, a non-linear $\mathcal{M_*}-\mathcal{M}_\mathrm{BH}$ relation has also been identified with observations on local galaxies \citep[e.g.,][]{graham16,sahu19}, where the steepened low-mass end was found to mainly consist of Sersic galaixes while the flattened massive end is dominated by the gas-poor core-Sersic galaxies with partially depleted cores caused by dry mergers. The absence of the $\mathcal{M_*}-\mathcal{M}_\mathrm{BH}$ steepening at the low-mass end suggests that the host galaxies of our sample are not undergoing such a transition phase. However, with the limited spatial resolution of our imaging data (Section \ref{sec:data}) that prevent us from investigating the galaxy morphology of our sample, we cannot rule out the possibility that the morphology dependent $\mathcal{M_*}-\mathcal{M}_\mathrm{BH}$ relation can already been identified at $z\sim2$.

\subsection{Impact of Model Assumptions and Data Limitations} \label{subsec:assumptions}
Although in Section \ref{subsec:systematics} we have incorporated possible systematics of from measurement uncertainties and selection biases, our results and discussions may still be impacted by several assumptions. Here we discuss several assumptions that may impact our results. 

Our first assumption is the SEDs of AGN. In our fitting we assume the AGN template of typical type 1 AGN and approximate the star forming history of host galaxies with a single exponential law. While such assumptions are sufficient for typical type 1 AGN, they may inevitably fail to describe some specific objects, resulting in the failure of fitting. We perform a simple test to check whether or not the SED fitting quality may introduce additional biases to our result, focusing on the selection criterion of $\chi^2_\mathrm{red}$ of the SED fitting results (Section \ref{subsec:mstar}). In addition to our original sample 66 objects selected with $\chi^2_\mathrm{red} < 5$, we define a subsample of 55 objects with $\chi^2_\mathrm{red} < 2$, then performing the same analysis as mentioned in Section \ref{sec:model}. The resulting intrinsic $\mathcal{M_*}-\mathcal{M}_\mathrm{BH}$ relation derived with this subsample would have a $0.43_{-0.19}^{+0.17}$ positive offset from the local relation, which is in agreement with the results derived with our full sample of 66 objects in Section \ref{sec:int_results}. The consistent results suggest that our original $\chi^2_\mathrm{red} < 5$ criterion would not introduce additional biases on the measurements of $\mathcal{M_*}-\mathcal{M}_\mathrm{BH}$ relation. 

Second, we examine the assumptions of active fraction in Section \ref{sec:model} from which we perform the forward modelling. When generating $\mathcal{M_*}$ and $\mathcal{M}_\mathrm{BH}$ of the mock intrinsic sample, we assume the distributions galaxies are the same as those of AGN, i.e., ignoring the $\mathcal{M_*}$ or $\mathcal{M}_\mathrm{BH}$ dependency of the galaxy active fraction. At high z, the active fraction can be derived by comparing BHMF to the total SMF as showed in \citet{schulze15}. They found type~1 AGN fraction at $z=2$ is nearly constant at $\log(\mathcal{M}_\mathrm{BH}/M_\odot) < 8.5$, before increasing towards the massive end up to $\log(\mathcal{M}_\mathrm{BH}/M_\odot) \sim 9.6$ and finally dropping again. However, the uncertainties at the massive end is large due to the limited number of objects and different assumptions of the functional form of BHMF \citet{schulze15}. Alternatively, \citet{aversa15} derive the ERDF analytically by fitting the continuity equation, finding the active fraction at the similar redshift monotonically increases with $\mathcal{M}_\mathrm{BH}$. To examine whether or not applying AGN active fraction would change our results, we assign a weight to each mock galaxy based on the active fraction given by the $\mathcal{M}_\mathrm{BH,true}$. We then randomly sample the mock galaxies with their weights, and perform the subsequent forward modelling procedure same as introduced in Section \ref{sec:model} to obtain the best-fit offset of the intrinsic relation. We find that applying the \citet{schulze15} and \citet{aversa15} active fraction would result in a positive offset of $+0.47_{-0.13}^{+0.16}$ and $+0.44_{-0.17}^{+0.14}$, respectively, for the best-fit intrinsic relation. Both of two results are slightly smaller but consistent with the offset we obtained in Section \ref{sec:int_results}. We also confirm that when applying these two different active fractions, the TNG non-linear relation would still be disfavored at 97\% level, suggesting that the impact of active fraction would not affect our conclusions.


Another assumption that may impact our results is the intrinsic scatter of the $\mathcal{M_*}-\mathcal{M}_\mathrm{BH}$ relation. It has been suggested that the redshift evolution of intrinsic scatter may imply the origin of the observed local $\mathcal{M_*}-\mathcal{M}_\mathrm{BH}$ relation. If the local relation were due to random mergers of galaxies, The intrinsic scatter would increase with z. Such a scenario is not supported by obervational results, as it has been found that the intrinsic scatter remains relatively constant out to $z\sim2$ with a value of $0.36\pm0.06$ \citep{ding20}. To test whether or not different assumptions of intrinsic scatter would affect our results, we repeat the forward modelling analysis in Section \ref{sec:model} with different intrinsic scatter values of (0.2, 0.3, 0.4). The corresponding offset of the intrinsic $\mathcal{M_*}-\mathcal{M}_\mathrm{BH}$ relation is ($+0.52\pm0.14, +0.52\pm0.14, +0.44_{-0.17}^{+0.19}$), indicating that our best-fit offset is stable when assuming different intrinsic scatter values suggested by previous studies. However, a larger sample with smaller measurement errors in $\mathcal{M_*}$ and $\mathcal{M}_\mathrm{BH}$ is in need to simultaneously fit the intrinsic scatter and offset. Future follow-up observations on the Mg{\sc ii} or H$\beta$ emission lines on our targets would reduce the uncertainty in  $\mathcal{M}_\mathrm{BH}$ measurements, helping to distinguish the evolutionary pattern of AGN with low $\mathcal{M}_\mathrm{BH}$ and low $\mathcal{M}_*$.





\section{Summary} \label{sec:summary}
We investigate the $\mathcal{M_*}-\mathcal{M}_\mathrm{BH}$ relation at $z=2.0-2.5$ with $66$ HETDEX type 1 AGN by measuring the $\mathcal{M}_\mathrm{BH}$ with the single-epoch virial method using the C{\sc iv} emission lines in the HETDEX spectra. The untargeted spectroscopy of the HETDEX survey allows identification of a type 1 AGN sample with BH masses down to $\log(\mathcal{M}_\mathrm{BH}/\mathcal{M}_\odot)\sim~7$, a mass range that has not been previously explored at this redshift. Based on multi-wavelength imaging data, we derive the $\mathcal{M_*}$ of our type 1 AGN by decomposing the stellar and nuclear light with SED fitting. Monte Carlo simulation confirms that our $\mathcal{M_*}$ estimation based on SED decomposition has no signs of bias at $\log(\mathcal{M}_*/\mathcal{M}_\odot) > 9.6$ with a systematic scatter of $\sim0.2$~dex. We also compare our $\mathcal{M_*}$ estimation with the image decomposition method based on the latest JWST ERS imaging data of two objects, finding that the consistency between the two methods for these two cases are within in 0.1 and 0.5~dex, respectively.

The forward modelling accounts for the observational biases, and allows derivation of the intrinsic $\mathcal{M_*}-\mathcal{M}_\mathrm{BH}$ relation. The intrinsic $\mathcal{M_*}-\mathcal{M}_\mathrm{BH}$ relation has a $0.52\pm0.14$~dex positive offset (logarithmic) from the local relation, suggesting a positive evolution of $\mathcal{M_*}-\mathcal{M}_\mathrm{BH}$ towards higher redshifts. This behavior may be caused by the gas-rich, compact nature of higher-redshift galaxies that would result in the more efficient accretion of gas to the central BHs.

Finally, 
the non-linear break towards low-$\mathcal{M}_\mathrm{BH}$ at the low mass regime predicted by the TNG100 hydrodynamic simulation is inconsistent with our observed relation at the $98\%$ confidence level. This result may be due to the SN feedback being weaker than predicted in the simulation.

\section{Acknowledgments}
HETDEX is led by the University of Texas at Austin McDonald Observatory and Department of Astronomy with participation from the Ludwig-Maximilians-Universit\"{a}t M\"{u}nchen, Max-Planck-Institut f\"{u}r Extraterrestrische Physik (MPE), Leibniz-Institut f\"{u}r Astrophysik Potsdam (AIP), Texas A\&M University, Pennsylvania State University, Institut f\"{u}r Astrophysik G\"{o}ttingen, The University of Oxford, Max-Planck-Institut f\"{u}r Astrophysik (MPA), The University of Tokyo and Missouri University of Science and Technology. In addition to Institutional support, HETDEX is funded by the National Science Foundation (grant AST-0926815), the State of Texas, the US Air Force (AFRL FA9451-04-2- 0355), and generous support from private individuals and foundations.

The observations were obtained with the Hobby-Eberly Telescope (HET), which is a joint project of the University of Texas at Austin, the Pennsylvania State University, Ludwig-Maximilians-Universit\"{a}t M\"{u}nchen, and Georg-August-Universit\"{a}t G\"{o}ttingen. The HET is named in honor of its principal benefactors, William P. Hobby and Robert E. Eberly.

VIRUS is a joint project of the University of Texas at Austin, Leibniz-Institut f\"{u}r Astrophysik Potsdam (AIP), Texas A\&M University, Max-Planck-Institut f\"{u}r Extraterrestrische Physik (MPE), Ludwig-Maximilians-Universit\"{a}t M\"{u}nchen, The University of Oxford, Pennsylvania State University, Institut f\"{u}r Astrophysik G\"{o}ttingen, and  Max-Planck-Institut f\"{u}r Astrophysik (MPA).

The authors acknowledge the Texas Advanced Computing Center\footnote{\url{http://www.tacc.utexas.edu}} (TACC) at The University of Texas at Austin for providing high performance computing, visualization, and storage resources that have contributed to the research results reported within this paper. 

This work is based on observations made with the NASA/ESA/CSA James Webb Space Telescope. The data were obtained from the Mikulski Archive for Space Telescopes at the Space
Telescope Science Institute, which is operated by the Association of Universities for Research in Astronomy, Inc., under NASA contract NAS 5-03127 for JWST. These observations are associated with program 1345. We acknowledge the CEERS team led by Steven L. Finkelstein for developing their observing programs with a zero-exclusive-access period.

The Hyper Suprime-Cam (HSC) collaboration includes the astronomical communities of Japan and Taiwan, and Princeton University. The HSC instrumentation and software were developed by the National Astronomical Observatory of Japan (NAOJ), the Kavli Institute for the Physics and Mathematics of the Universe (Kavli IPMU), the University of Tokyo, the High Energy Accelerator Research Organization (KEK), the Academia Sinica Institute for Astronomy and Astrophysics in Taiwan (ASIAA), and Princeton University. Funding was contributed by the FIRST program from the Japanese Cabinet Office, the Ministry of Education, Culture, Sports, Science and Technology (MEXT), the Japan Society for the Promotion of Science (JSPS), Japan Science and Technology Agency (JST), the Toray Science Foundation, NAOJ, Kavli IPMU, KEK, ASIAA, and Princeton University. 

This paper makes use of software developed for the Large Synoptic Survey Telescope. We thank the LSST Project for making their code available as free software at \url{http://dm.lsst.org}.

This research is based in part on data collected at Subaru Telescope, which is operated by the National Astronomical Observatory of Japan. We are honored and grateful for the opportunity of observing the Universe from Maunakea, which has the cultural, historical and natural significance in Hawaii.

We fully appreciate the valuable comments from the anonymous referee that improved the clarity of this manuscript. Y.Z. thanks John D. Silverman and Niv Drory for their inputs during the completion of this work. This work is supported by the World Premier International Research Center Initiative (WPI Initiative), MEXT, Japan, as
well as KAKENHI Grant-in-Aid for Scientific Research (A) (20H00180, and 21H04467) through the Japan Society for the Promotion of Science (JSPS). Y.Z. acknouledges the support from JST SPRING (JPMJSP2108), as well as the joint research program of the Institute for Cosmic Ray Research (ICRR), University of Tokyo. 



\vspace{5mm}

\software{Astropy \citep{astropy13}, CIGALE \citep{boq19}, GALFIT \citep{peng02,peng10}, Source Extractor \citep{BA96}
          }
\bibliography{bib.bib}{}



\end{document}